\documentclass[
               dvips, usenatbib, fleqn
               ]{mn2e}

\usepackage{mathptmx}
\usepackage[dvips]{graphicx}
\usepackage{amsmath}
\usepackage{mn2e_fixbib}

\renewcommand{\bmath}[1]{\pmb{#1}}

\begin{document}

\title[Planetary nebulae from low-mass red giants]{%
  Planetary nebulae after common-envelope phases initiated by low-mass red giants}

\author[P.~D.~Hall et al.]{%
  Philip~D.~Hall,$^1$\thanks{E-mail:~pdh37@ast.cam.ac.uk}  Christopher~A.~Tout,$^1$ Robert~G.~Izzard$^2$ and Denise~Keller$^{2,3}$\\
  $^1$University of Cambridge, Institute of Astronomy, The Observatories, Madingley Road, Cambridge CB3 0HA \\
  $^2$Argelander-Institut f\"{u}r Astronomie, University of Bonn, Auf dem H\"{u}gel 71, D-53121 Bonn, Germany \\
  $^3$Max-Planck-Institut f\"{u}r Radioastronomie, Auf dem H\"{u}gel 69, D-53121
Bonn, Germany
  }
\date{Accepted 2013 July 30. Received 2013 July 30; in original form 2013 June 28}
\volume{435}
\pagerange{2048--2059}
\pubyear{2013}
\maketitle

\label{firstpage}

\begin{abstract}
It is likely that at least some planetary nebulae are composed of matter which was ejected from a binary star system during common-envelope (CE) evolution.
For these planetary nebulae the ionizing component is the hot and luminous remnant of a giant which had its envelope ejected by a companion in the process of spiralling-in to its current short-period orbit.
A large fraction of CE phases which end with ejection of the envelope are thought to be initiated by low-mass red giants, giants with inert, degenerate helium cores.
We discuss the possible end-of-CE structures of such stars and their subsequent evolution to investigate for which structures planetary nebulae are formed.
We assume that a planetary nebula forms if the remnant reaches an effective temperature greater than $30\,\rmn{kK}$ within $10^4\,\rmn{yr}$ of ejecting its envelope.
We assume that the composition profile is unchanged during the CE phase so that possible remnant structures are parametrized by the end-of-CE core mass, envelope mass and entropy profile.
We find that planetary nebulae are expected in post-CE systems with core masses greater than about $0.3\,\rmn{M}_{\sun}$ if remnants end the CE phase in thermal equilibrium.
We show that whether the remnant undergoes a pre-white dwarf plateau phase depends on the prescribed end-of-CE envelope mass.
Thus, observing a young post-CE system would constrain the end-of-CE envelope mass and post-CE evolution.
\end{abstract}


\section{Introduction}
\label{sec:intro}
Planetary nebulae (PNe) are large expanding shells of gas which are visible because they are photoionized by a hot and luminous central star from which they were ejected \citep{Kwo00:0}.
Traditionally they are explained as having formed from the envelopes of isolated low- and intermediate-mass asymptotic giant branch (AGB) stars.
At the end of the AGB phase, stars have strong winds which cause them to lose most of their envelopes and begin to contract to become carbon--oxygen-core white dwarfs.
Fast winds during the contraction phase interact with the previously ejected circumstellar matter to form a PN, ionized by the hot remnant core.

However, there are more than $40$ known PNe for which the compact source of ionizing photons is a member of a short-period binary star system.
This sub-class could constitute $10$--$20$ per cent of the total PNe population \citep{DeM09:316, Mis09:813}.
These PNe may have formed by a different channel, after a common-envelope (CE) phase which began when a giant star overfilled its Roche lobe in a longer period orbit \citep{Pac76:73}.
In such a situation the mass-loss rate from the giant rapidly increases to a rate at which the companion is unable to accrete and the companion becomes immersed in the common giant envelope and spirals towards the giant's core, depositing energy and angular momentum in the envelope.
This continues until the components merge to form a single star or until the envelope is ejected, to leave the remnant of the giant and its companion in a shorter period binary system. 
In this way, the material necessary for the PN (the ejected envelope) and the short-period binary system which may be able to illuminate the envelope are arranged.
Despite its importance, our knowledge of CE evolution remains uncertain.

A CE phase can be classified by the structure of the mass-losing star at the beginning of the phase and whether the envelope is ejected or it ends with the merging of the two components.
The structure of the loser dictates whether a CE phase begins and, when the envelope is ejected, the type of remnant.
The initial rate of mass loss, and thus whether a CE phase begins when a star fills its Roche lobe, is dictated by the response of its radius and Roche lobe to mass loss. 
A simple analysis of these responses implies that a CE phase can begin after the onset of Roche lobe overflow of stars with deep convective envelopes \citep{Web85:39}. 
Low- and intermediate-mass stars have such a structure and a developed core on their first ascent of the giant branch as red giant branch (RGB) stars and their second as AGB stars.
In synthetic stellar populations the majority of CE phases ending with successful ejection of the envelope are initiated by stars of these types.
This means that PNe are expected to form from the envelopes of both RGB and AGB stars via the CE channel.
In the traditional single-star channel only AGB stars produce PNe, because the wind of an isolated star is not thought to be strong enough to remove its envelope during the RGB phase.
It is thus interesting to consider the question of how important CE phases initiated by RGB stars are as a formation channel for PNe.
Here, we focus on low-mass RGB stars, also known as low-mass red giants.
These are stars with inert electron-degenerate helium cores, thin hydrogen burning shells and large convective envelopes and have zero-age masses less than about $2\,\rmn{M}_{\sun}$.

There are many observed short-period binary systems containing low-mass ($M \la 0.5\,\rmn{M}_{\sun}$) white dwarfs, which could have formed after CE phases initiated by low-mass RGB stars \citep{Zor11:L3}.
However these are observed long after the ejection of the envelope.
We would like to know whether these systems had a PN phase in their past.
Synthetic populations can be used to predict the number of PNe formed after CE phases initiated by low-mass red giants if it is possible to predict whether the ejected envelope becomes a PN, and the lifetime of its PN phase, on the basis of its pre- or post-CE properties \citetext{\citealp{deK90:189}; \citealp*{Yun93:794}; \citealp{Moe06:916}}.

Whether or not the ejected envelope becomes a PN depends on both the properties of and interaction between the envelope and the central remnant binary system.
However, an important necessary condition is that the remnant supplies the ejected envelope with ionizing photons at a sufficient rate and for a sufficient period of time before the expanding envelope becomes too tenuous to appear as a PN.
\citet{Ibe89:505, Ibe93:343} and \citet{Yun93:794}, for example, gave the condition that a CE remnant must achieve an effective temperature greater than $T_{\rmn{eff,crit}} = 30\,\rmn{kK}$ within $10^4\,\rmn{yr}$ of the end of the CE phase.
This simple condition is based on observations of PNe: observed central stars have effective temperatures greater than the critical value and estimated nebular lifetimes, based on their size and expansion speeds, of about $10^4\,\rmn{yr}$ \citep{Kwo00:0}.

The proposed condition is simple enough to be checked in the evolution of a model CE remnant without knowing the evolution of the ejected envelope.
However, we must compute evolutionary models of remnants and relate the properties of the system at the beginning and end of the CE phase.
For now, there are no realistic models of CE evolution from start to finish: the most recent three-dimensional hydrodynamical models of CE evolution end after an initially rapid phase of spiral-in and before most of the envelope is unbound \citep{Ric12:74, Pas12:52}.
There is uncertainty about what happens after this \citep{Kas11:1466} so it is not possible to predict the structure of the remnant of the giant, its composition profile, entropy profile and mass, from the outputs of such models.
Instead it is usual to assume a prescription to predict the orbital properties and transform the pre-CE giant into its post-CE counterpart.
At the end of the CE phase the remnant has the same composition profile, and therefore the same core mass, as its pre-CE counterpart but with a sufficiently small envelope mass that it contracts and becomes a white dwarf.
That the composition profile is unchanged during the phase is based on the reasonable assumption that the duration of a CE phase is much shorter than the time-scale for core growth.
It is also necessary to account for mass loss by wind during post-CE evolution.
This is because the remnants which sustain hydrogen-burning shells evolve as a sequence of approximately quasi-static, equilibrium burning models in which the small envelope mass is depleted by hydrogen burning at its base and a wind.
The rate at which the envelope mass decreases dictates the time-scale on which the remnant evolves to high effective temperatures and thus whether it satisfies the condition for the formation of a PN.
Hence we need to prescribe the relation between the pre- or post-CE system properties to the end-of-CE structure of the remnant as well as the post-CE wind mass loss if we are to check for the formation of a PN in a particular case.

\citet{Ibe89:505, Ibe93:343} have also considered post-RGB PNe as part of a wider discussion of PNe formed after CE phases.
They concluded that a PN is not formed after a CE phase initiated by a low-mass RGB star unless the remnant has a core mass greater than $0.4\,\rmn{M}_{\sun}$ and ends the CE phase inside a Roche lobe of radius less than about $1\,\rmn{R}_{\sun}$ \citep{Ibe95:2}.
They found that remnants with core masses less than about $0.3\,\rmn{M}_{\sun}$ do not achieve sufficiently high effective temperatures.
Remnants which have core masses greater than this must have end-of-CE Roche lobe radii less than $1\,\rmn{R}_{\sun}$ if they are to reach the critical effective temperature rapidly enough to avoid being lazy.
In a typical model population, this implies that some PNe are formed from this channel \citep{Yun93:794}.
Although no observed PNe have been clearly identified as post-RGB objects \citep{DeM09:316}, there are some possible systems (see Section~\ref{sec:discussion}).
Some studies assume that post-RGB systems evolve too slowly to form PNe \citep*{Nie12:2764}.
However, the post-CE evolution depends on the uncertain end-of-CE structure. 
We would like to know if we expect PNe after CE phases initiated by low-mass RGB stars because they would allow us to learn more about CE evolution and the formation of PNe.

Our main contribution to the question of the importance of CE phases initiated by low-mass RGB stars as a PNe formation channel is to consider alternative prescriptions for the end-of-CE structure.
We also extend the work of \citet{Ibe89:505, Ibe93:343} by computing a larger set of models, by including stars with zero-age metallicity $Z_0=0.001$ as well as those with $Z_0=0.02$ and by considering the effect of a post-CE wind.
In Section~\ref{sec:method} we discuss possible end-of-CE structures and argue that only thermal equilibrium remnants are expected to form PNe.
We describe our stellar-evolution code and our method of modelling remnants of this type.
In Section~\ref{sec:results} we discuss the results.
We derive regions in the end-of-CE phase core mass--envelope mass plane for which the formation of a PN is possible and discuss how these regions change with the uncertain post-CE wind.
In Section~\ref{sec:discussion} we discuss the implications for the significance of CE phases initiated by low-mass red giants as a PN formation channel and conclude in Section \ref{sec:conclusions}.

\section{Method}
\label{sec:method}
We start by describing the conditions required for the formation of a PN and then discuss proposed prescriptions for the end-of-CE structure and orbital properties.

\subsection{Conditions for planetary nebula formation}
\label{sec:PNconditions}
In Section~\ref{sec:intro} we gave a condition for PN formation, that remnants reach an effective temperature greater than $T_{\rmn{eff,crit}} = 30\,\rmn{kK}$ within $10^4\,\rmn{yr}$ of the end of the CE phase.
This defines the central star of planetary nebula (CSPN) region: the region of the theoretical Hertzsprung--Russell (HR) diagram for which $T_{\rmn{eff}} \geq T_{\rmn{eff,crit}}$.
It is also useful to define the transition time $t_{\rmn{trans}}$ to be the time between the end of the CE phase and the star first entering the CSPN region.
Our condition for PN formation is based on observations of PNe: observed central stars have effective temperatures and luminosities within the CSPN region and estimated nebula lifetimes, based on their size and expansion speeds, of about $10^4\,\rmn{yr}$ \citep{Kwo00:0}.
We define the fading time $t_{\rmn{fade}}$ to be the time spent passing through CSPN region.
The calculation of this quantity allows us to check that those systems which pass through the CSPN region also spend sufficient time there to be observed.

\subsection{Prescriptions for the end-of-CE structure}
A problem is to find the configuration of the system at the end of an assumed CE phase, given the structure of a binary system at the beginning.
We want to determine whether the envelope is successfully ejected and, if it is, the orbital configuration and structure of the giant's remnant at the end of the CE phase.
There are two aspects of the end-of-CE structure to be determined, the envelope mass and entropy profile.
We assume that the composition profile is unchanged by the CE phase.
We also require that, to be a valid end-of-CE structure, the star must only contract during the post-CE phase so that the system is detached and the remnant does not immediately interact with its companion.
Remnant models can be produced using one-dimensional stellar-evolution codes and by removing mass from models of pre-CE giants.

The remnants of low-mass red giants computed by \citet{Ibe86:742, Ibe93:343} were in thermal equilibrium with the end-of-CE envelope mass fixed by requiring the radius to equal the Roche lobe radius.
They computed the evolution of the orbit and Roche lobe radius by assuming an efficiency for the loss of orbital energy in ejecting the envelope.
Such remnants undergo a plateau phase of hydrogen burning along a roughly horizontal track in the theoretical HR diagram (Fig.~\ref{fig:TeffL1}).
These thermal equilibrium, contracting remnants are only possible for a range of envelope masses at any given core mass.

We define the peel-off envelope mass to be the maximum envelope mass of such remnants.
This is the envelope mass for a given core mass such that the star contracts and evolves away from the giant branch.
It has been discussed previously \citetext{\citealp{Ref70:426}; \citealp*{Cas92:227}}.
The peel-off envelope mass is about $10^{-2}\,\rmn{M}_{\sun}$ for both $Z_0=0.02$ and $Z_0=0.001$ stars.

We define the knee envelope mass to be the minimum envelope mass of thermal equilibrium remnants.
\citet{Dei70:671} found a difference in the evolution of adiabatic mass-loss remnants with envelope masses greater and less than the knee value, although they did not refer to it in these terms.
For envelope masses greater than the knee value, the remnants regain thermal equilibrium to begin a plateau phase of hydrogen burning.
For envelope masses less than the knee value, the remnants fail to regain thermal equilibrium and evolve directly to become white dwarfs.

\citet{Ge10:724} and \citet{Del10:L28} computed adiabatic mass-loss sequences, fixing the entropy profile in the remnant to be the same as that in the pre-CE giant.
This is the approximate limit of very fast mass loss which could be appropriate to model the rapid ejection of the envelope.
They assumed the rapid ejection phase ends when the star contracts inside its Roche lobe.
They computed the evolution of the orbit and Roche lobe radius by assuming an efficiency for the loss of orbital energy in ejecting the envelope.
In their prescription the remnant ends the ejection phase in thermal non-equilibrium with the entropy profile of the giant and small envelope mass.
This is probably not the end of the CE phase, as we define it here, because such remnants can expand in an attempt to regain thermal equilibrium and further interact with the companion.
\citet{Iva11:76} considered the effect of this thermal readjustment phase although, rather than removing mass adiabatically, she modelled the ejection by applying very rapid mass-loss rates (about $1\,\rmn{M}_{\sun}\,\rmn{yr}^{-1}$) which she referred to as semi-adiabatic mass loss.
She investigated the post-ejection thermal readjustment by evolving remnants with different end-of-ejection envelope masses.
She found a critical envelope mass which she identified with the maximum of the ratio of pressure to density ($P/\rho$), or equivalently the maximum of the sound speed, in the hydrogen burning shell of the pre-CE giant.
This point has been called the divergence point.
If the end-of-ejection mass of the adiabatic remnant is greater than the divergence point mass it expands in an attempt to regain thermal equilibrium.
She argued that if the ejection phase ends with a greater envelope mass than the divergence point value, the remaining mass is lost in a stable Roche-lobe overflow phase or leads to the merging of the components.
Therefore, in this prescription, the end-of-CE remnant has a thermal non-equilibrium entropy profile with the divergence point envelope mass.

Others have suggested different prescriptions for the end-of-CE envelope mass but without explicitly discussing the entropy profile or post-CE evolution.
These prescriptions are motivated by the need to find the base of the ejected envelope for the $\alpha \lambda$ scheme \citep*{Hur02:897}.
In this scheme, $\alpha$ is the ratio of the change in orbital energy to the binding energy of the ejected envelope which is used to infer the properties of the system at the end of the CE phase \citep{DeM11:2277, Iva13:59}.
The inner boundary of the ejected envelope, the remnant--envelope boundary, is required to evaluate the binding energy of the remnant and its envelope, parametrized by $\lambda$.
\citet{Tau01:170} and \citet{Iva11:91} summarized suggestions for this remnant--envelope boundary, including (i) the edge of the helium core, the outermost point where the hydrogen mass fraction $X=0$, (ii) the point where the hydrogen mass fraction $X=0.1$, (iii) the point of maximum energy generation ($\varepsilon_{\rmn{nuc}}$) plus $10^{-3}\,\rmn{M}_{\sun}$ \citep{DeM11:2277} and (iv) the base of the convective envelope \citep{Tau01:170}.
There is not much justification for these prescriptions but they provide simply identifiable points for computing the binding energy at the beginning of a CE phase.
They all imply that end-of-CE envelope masses are a function of core mass only.
We refer to these as simple prescriptions.

To summarize, for remnants which end the CE phase in thermal equilibrium, there are a range of possible end-of-CE masses between the knee and peel-off masses.
For remnants which end the CE phase in a thermal non-equilibrium configuration, in the case in which the entropy profile is fixed in a sequence of semi-adiabatic mass loss, the envelope mass must be less than or equal to the divergence point identified by \citet{Iva11:76} because it is only for these envelope masses that the remnant contracts.
Table~\ref{tab:masses} is a summary of the critical masses we have defined.
We now discuss the evolution of remnants of the two types, thermal equilibrium and thermal non-equilibrium to see whether they can form PNe.
We also compute models of thermal equilibrium remnants to extend the work of \citet{Ibe89:505}.
We discuss the simple prescriptions only for these two possibilities for the entropy profile.

\begin{table*}
\begin{minipage}{11.2cm}
  \centering
  \caption{Summary of critical masses for a mass-loss remnant of a low-mass red giant of given core mass.
Each can be referred to in terms of a post-mass loss envelope mass and as a point in the pre-mass loss red giant.}
  \label{tab:masses}
  \begin{tabular}{@{}p{2.8cm}p{8.4cm}}
   \hline
   Critical mass & Description  \\
   \hline
   Peel-off mass & The peel-off mass is the maximum mass of a contracting thermal equilibrium remnant.
A thermal equilibrium remnant with mass greater than this is a red giant and expands as its core mass grows. 
The peel-off envelope mass is about $10^{-2}\,\rmn{M}_{\sun}$ for a $0.3540\,\rmn{M}_{\sun}$ core with $Z_0=0.02$. \\
   Knee mass & The knee mass is the minimum mass of a thermal equilibrium remnant. 
A thermal equilibrium remnant with mass greater than this but less than the peel-off mass undergoes a plateau phase until the envelope mass decreases to the knee value, the burning shell fades, and the star becomes a helium-core white dwarf. 
The knee envelope mass is $1.2 \times 10^{-3}\,\rmn{M}_{\sun}$ for a $0.3540\,\rmn{M}_{\sun}$ core with $Z_0=0.02$. \\
   Divergence point mass & The divergence point is at the maximum of the ratio of pressure to density inside the thin hydrogen-burning shell of the giant. 
The divergence point mass is the mass enclosed at this point.
\citet{Iva11:76} found that semi-adiabatic mass-loss remnants with masses less than this contract, and those with masses greater than this expand.
She suggested this is the end-of-CE mass.
The divergence point envelope mass is $0.3 \times 10^{-3}\,\rmn{M}_{\sun}$ for a $0.3540\,\rmn{M}_{\sun}$ core with $Z_0=0.02$. \\
   \hline
 \end{tabular}
\end{minipage}
\end{table*}

\subsection{The Cambridge stellar-evolution code}
We use the version of the Cambridge stellar-evolution code described by \citet{Sta09:1699} which has descended from that written by \citet{Egg71:351, Egg72:361} and was updated by \citet{Pol95:964}. 
Model sequences produced by the code satisfy a standard set of one-dimensional quasi-static stellar-evolution equations with meshpoints distributed in a non-Lagrangian mesh \citep{Egg71:351}. 
The convective mixing-length \citep{Boh58:108} parameter $\alpha_{\rmn{MLT}}=2$ and convective overshooting is included with a parameter $\delta_{\rmn{ov}}=0.12$ \citep*{Sch97:696} for consistency with the work of \citet*{Hur00:543}, whose analytic formulae for stellar evolution can be used for population synthesis. 
These convective parameters are consistent with observations of the Sun, open clusters and spectroscopic binary systems \citep{Pol98:525}. 
At the zero-age main sequence (ZAMS) the models are in complete thermal equilibrium with uniform abundance profiles.
We choose the zero-age helium mass fraction $Y_0=0.24+2Z_0$, so that there is a constant rate of helium to metal enhancement, $\Delta Y/\Delta Z=2$  from the primordial abundance $(Y_{\rmn{p}}, Z_{\rmn{p}})=(0.24, 0)$, calibrated to the solar-like abundance $(0.28, 0.02)$. 
For $Z_0=0.001$ this gives $Y_0=0.242$.
The relative abundances of the various metals in the initial configuration are constant throughout all models, scaled to the solar abundances given by \citet{And89:197}.
In our models only the mass fractions of ${}^1\rmn{H}$, ${}^3\rmn{He}$, ${}^4\rmn{He}$, ${}^{12}\rmn{C}$, ${}^{14}\rmn{N}$, ${}^{16}\rmn{O}$ and ${}^{20}\rmn{Ne}$ are evolved by convective mixing and nuclear reactions.
These nuclides are sufficient to determine the structure of the stars of interest here.

The equation of state and other thermodynamic quantities are described by \citet*{Egg73:325} and \citet{Pol95:964}. 
The radiative opacity is that of the OPAL collaboration \citep{Igl96:943}, supplemented by the molecular opacities of \citet{Ale94:879} and \citet{Fer05:585} for low temperatures and of \citet{Buc76:440} for pure electron-scattering at high temperatures. 
The electron conduction opacity is from the work of \citet{Hub69:18} and \citet{Can70:641}. 
The construction of the opacity tables and their inclusion in the code was described by \citet{Eld04:201}. 
The nuclear-reaction rates are those of \citet{Cau88:283} and the NACRE collaboration \citep{Ang99:3}, as described by \citet{Eld04:87} and \citet{Sta05:375}. 
The enhancement of reaction rates by electron screening is included according to \citet{Grab73:457}.
The rates of energy loss in neutrinos are due to \citet{Ito89:354, Ito92:622} for the photo/pair and plasma processes, respectively and \citet{Ito83:858} and \citet*{Mun87:708} for the bremsstrahlung process.

\subsection{Thermal equilibrium remnants}
\label{sec:mdotphase}
To make a thermal equilibrium remnant we first evolve a star of zero-age metallicity $Z_0$ and low mass $M_0$ at constant mass until it becomes a red giant with a chosen helium core mass $M_{\rmn{core}}$, where the outer boundary of the helium core is defined to be the point at which the mass fraction of hydrogen $X=0.1$.
We then apply a scaled Reimers' mass loss \citep{Rei75:369},
\begin{equation}
  \frac{\rmn{d}M}{\rmn{d}t} = 
  - 4 \times 10^{-13} \eta
  \left(\frac{R}{\rmn{R}_{\sun}}\right) \left(\frac{L}{\rmn{L}_{\sun}}\right) \left(\frac{M}{\rmn{M}_{\sun}}\right)^{-1}
  \,\rmn{M}_{\sun}\,\rmn{yr}^{-1},
\label{eq:Reimersrate}
\end{equation}
where $t$ is time, $R$ the radius, $L$ the luminosity and $M$ the mass of the star. 
We artificially set the parameter $\eta$ to be sufficiently large that the star maintains an approximately constant core mass during the mass-loss phase but is not significantly driven out of thermal equilibrium.
We find that $\eta=500$ works well for all cases.
When the star's envelope mass has decreased to a chosen final value, we switch off mass loss and allow the remnant to evolve with constant total mass.
We choose the final envelope mass to be sufficiently small that the star contracts and becomes a white dwarf.
That is, the envelope mass is less than the peel-off envelope mass.
During the subsequent post-CE evolution, the envelope mass decreases because of the hydrogen-burning shell at its base and the star contracts to become a helium-core white dwarf composed of an inert degenerate helium core surrounded by a fading hydrogen-burning shell and a low-mass hydrogen-rich envelope.
The model is evolved for sufficiently long after the end of the CE phase to check whether it satisfies the conditions for PN formation.

\begin{figure}
  \includegraphics[width=84mm]{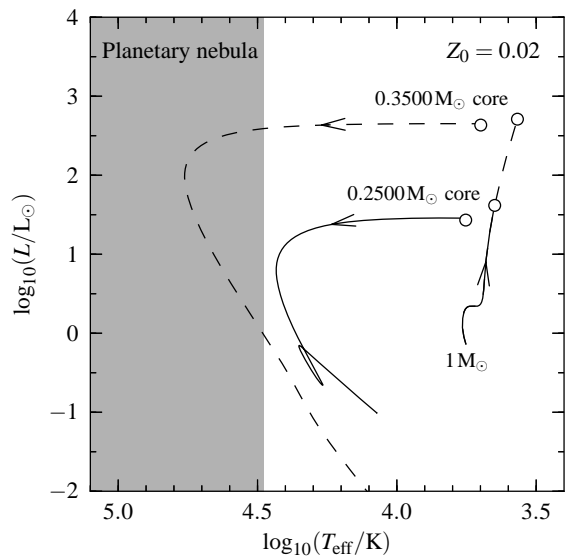}
  \caption{
Evolution of two $Z_0=0.02$ models in the HR diagram during their pre- and post-CE phases. 
On the right-hand side the stars evolve from the ZAMS until they become low-mass red giants with core masses of $0.2500$ and $0.3500\,\rmn{M}_{\sun}$. 
We do not plot the artificial mass-loss phase.
On the left-hand side the stars emerge from the CE phase. 
Thin shell hydrogen burning continues so that the star represented by the dashed line has an end-of-CE core mass of $0.3505\,\rmn{M}_{\sun}$. 
For the solid line the end-of-CE core mass is $0.2531\,\rmn{M}_{\sun}$. 
Our CSPN/PN region is shaded.}
  \label{fig:TeffL1}
\end{figure}

\begin{figure}
  \includegraphics[width=84mm]{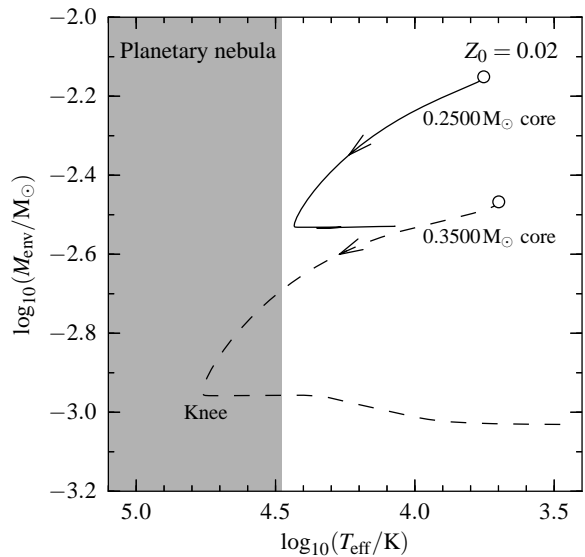}
  \caption{
Evolution in the effective temperature--envelope mass plane during the post-CE phase of the two models shown in Fig.~\ref{fig:TeffL1}.
For both stars the envelope mass decreases because of hydrogen shell burning during the plateau phase.
The effective temperature of both stars increases until they reach the point of maximum effective temperature, the knee.}
  \label{fig:lgTeff-lgMenv}
\end{figure}

An example of a model made in this way has zero-age metallicity $Z_0=0.02$ and mass $M_0=1\,\rmn{M}_{\sun}$.
We evolve this star from the ZAMS until it becomes a red giant with $M_{\rmn{core}}=0.3500\,\rmn{M}_{\sun}$, a radius of $55.4\,\rmn{R}_{\sun}$ and luminosity of $512\,\rmn{L}_{\sun}$.
Its evolution in the HR diagram is shown in \mbox{Fig.~\ref{fig:TeffL1}}. 
At this stage we remove mass with $\eta=500$.
During the artificial mass-loss phase, the star loses mass at a rate of about $10^{-5}\,\rmn{M}_{\sun}\,\rmn{yr}^{-1}$ and after some initial expansion the envelope contracts.
Global thermal equilibrium is barely disturbed during this phase.
Mass loss is stopped when the star has contracted to half its initial radius, $R = 27.7\,\rmn{R}_{\sun}$, to produce a thermal equilibrium remnant.
This radius is larger than is usually expected for the end of a CE phase, but the plateau evolution is such that stopping mass loss at high radius effectively allows us to investigate end-of-CE remnants with smaller envelope masses and radii (Section~\ref{sec:paramcomb}).
Thus we can derive, for a given core mass, a maximum end-of-CE envelope mass or radius for the formation of a PN.
At the point at which mass loss ends the envelope mass is $3.4\times 10^{-3}\,\rmn{M}_{\sun}$, the luminosity is $428\,\rmn{L}_{\sun}$, the core mass is $0.3505\,\rmn{M}_{\sun}$ and the total mass is $0.3540\,\rmn{M}_{\sun}$.
At this core mass the knee envelope mass is $1.2\times 10^{-3}\,\rmn{M}_{\sun}$ and the divergence point envelope mass is about $0.3\times 10^{-3}\,\rmn{M}_{\sun}$.
From this point, the star evolves at constant mass and the envelope contracts with continued hydrogen burning in a shell at its base.
Its evolution in the effective temperature--envelope mass plane is shown in \mbox{Fig.~\ref{fig:lgTeff-lgMenv}}.
At first, the luminosity is approximately constant and well approximated by the luminosity of a giant with the same core mass.
This plateau phase ends when the star reaches the knee at a maximum effective temperature $57.8\,\rmn{kK}$.
Its luminosity has decreased to $96.9\,\rmn{L}_{\sun}$.
By this stage the envelope mass has fallen to $1.1\times 10^{-3}\,\rmn{M}_{\sun}$ and the star begins to become a helium-core white dwarf as its burning shell fades.
In \mbox{Fig.~\ref{fig:TeffL1}}, we also show the evolution of a remnant which was produced by removing mass from a giant with a $0.2500\,\rmn{M}_{\sun}$ core.
In this case, mass is removed at about $10^{-7}\,\rmn{M}_{\sun}\,\rmn{yr}^{-1}$ until the star contracts to half its initial radius, $5.42\,\rmn{R}_{\sun}$, the core mass is $0.2531\,\rmn{M}_{\sun}$ and the envelope mass is $7.1 \times 10^{-3}\,\rmn{M}_{\sun}$.
At this core mass the knee envelope mass is $3.2\times 10^{-3}\,\rmn{M}_{\sun}$ and the divergence point envelope mass is about $0.7\times 10^{-3}\,\rmn{M}_{\sun}$.
It also evolves through a plateau phase, reaching a maximum effective temperature of $27.1\,\rmn{kK}$ at the knee.
Its evolution ends at the onset of a hydrogen shell flash.
Neither of these two models forms a PN.
The $0.25\,\rmn{M}_{\sun}$ core fails to reach the critical effective temperature.
The $0.35\,\rmn{M}_{\sun}$ core takes $2.1 \times 10^5\,\rmn{yr}$ to reach the critical region, longer than the estimated $10^4\,\rmn{yr}$ lifetime of the nebula.

\subsection{Flashes}
\label{sec:flashes}
The evolution of the $0.25\,\rmn{M}_{\sun}$ model ends at the onset of a hydrogen-shell flash.
We should consider if such shell flashes affect whether a remnant forms a PN.
Hydrogen-shell flashes near the surface of helium-core white dwarfs have been discussed extensively in the literature \citetext{e.g., \citealp{Dri99:89}; \citealp*{Alt01:617}}.
These occur as the star begins descending the white dwarf cooling track.
It is not clear for what range of core masses hydrogen-shell flashes are expected, but here, like \citet{Dri98:123, Dri99:89} we find that they occur for core masses between $0.2$ and $0.3\,\rmn{M}_{\sun}$ for $Z_0=0.02$.
However, others have found that when diffusion is included flashes occur between $0.18$ and $0.41\,\rmn{M}_{\sun}$ \citep{Alt01:617}.
Irrespective of the range of core masses for which they occur, the onset of these flashes always begins around or below $L \approx 1\,\rmn{L}_{\sun}$ which, for most thermal equilibrium remnants, is more than $10^4\,\rmn{yr}$ after the end of the CE phase.
During the flash the stars only enter the CSPN region after about $10\,\rmn{Myr}$, too late to illuminate a PN.

Another phenomenon found in previous studies of stars of this type is the possibility of igniting helium in the electron-degenerate core in an early or late hot flash, during the plateau or white dwarf cooling phase, respectively \citep{Cas93:649}.
In a late hot helium-core flash, the star ignites helium after the knee stage, during the white dwarf cooling phase and it may undergo mixing between its core and envelope before becoming a hot subdwarf or extreme horizontal branch star, depending on the degree of mixing between the core and envelope \citep{Swei97:L23}.
This is again not important here because, like the hydrogen-shell flashes discussed in the previous paragraph, the late hot flashes typically begin at low luminosity on the cooling curve more than $10^4\,\rmn{yr}$ after the end of the CE phase.
A star could also undergo an early hot helium-core flash in which helium is ignited during the plateau phase.
This could affect a remnant which would otherwise become a PN by truncating its plateau evolution before it reaches the CSPN region.
This also leads the star to become a hot subdwarf which may be hot enough to ionize a PN but again the time to evolve to that stage is more than $10^4\,\rmn{yr}$.
\citet{Han02:449} computed the core mass required at the beginning of the CE phase for the star to ignite helium in the post-CE phase although they did not distinguish between early and late hot flashes.
This means that there is a small range of end-of-CE core mass below that at the tip of the red giant branch for which remnants which would satisfy the conditions for a PN do not because they undergo an early hot flash.

\subsection{Parameter combinations}
\label{sec:paramcomb}
Our method to produce thermal equilibrium mass-loss remnants is similar to that of \citet{Ibe86:742, Ibe93:343} and \citet{Dri98:123}, who also removed mass from low-mass red giants at high rates and significantly decreased the mass-loss rate while the stars still have large radii.
Both these studies used a constant mass-loss rate appropriately scaled for different core masses.
The specific form of the mass-loss rate should not matter provided the core does not grow significantly and its thermal equilibrium is not disturbed.
We do not find significantly different results if we vary the value of $\eta$ within these constraints.
It is convenient to use the scaled Reimers mass-loss formula because it automatically adjusts the mass-loss rate to the properties of giants of different core masses.

The two models discussed do not form PNe, but we would like to know for which combinations of end-of-CE core and envelope mass the formation of a PN is possible.
We vary the zero-age metallicity and the core mass.
We choose two values for the metallicity, $Z_0=0.02$ and $Z_0=0.001$.
For each of these we choose a single zero-age mass, $M_0=1\,\rmn{M}_{\sun}$, for the pre-CE evolution.
We only consider one zero-age total mass because the structure of the core of a low-mass red-giant depends only on the core mass \citep{Egg68:387}.
We choose a set of core masses between the minimum core mass, at the beginning of the red-giant phase, and the maximum core mass, at the beginning of the helium-core flash.
We take the beginning of the red-giant phase to be the stage at which the convective envelope first dredges down more than $2/5$ of the envelope mass \citep{Pol98:525}.  
For both metallicities, the core mass during the red-giant phase ranges from about $0.15\,\rmn{M}_{\sun}$ to $0.47\,\rmn{M}_{\sun}$ and we model a set of 13 cores in this range, in steps of $0.025\,\rmn{M}_{\sun}$ at the beginning of mass loss.

We efficiently investigate all end-of-CE envelope masses between the knee and peel-off.
We choose to end mass loss when the star shrinks to half its pre-mass loss radius, which in all cases gives envelope masses below the peel-off mass.
In stars with core mass greater than about $0.2\,\rmn{M}_{\sun}$ the composition profile is unaffected by the artificial mass-loss phase.
During the subsequent evolution the composition profile keeps the same shape but grows outward in mass by a small amount until the star reaches the knee.
Because of this property of the plateau evolution, by stopping mass loss when the mass has fallen below the peel-off mass, we also effectively investigate the evolution of stars which end the CE phase with envelope masses between this point and up to the knee.
In stars with core mass less than about $0.2\,\rmn{M}_{\sun}$ the composition profile changes during the artificial mass-loss phase.
This is because our $Z_0=0.02$, $M_0=1\,\rmn{M}_{\sun}$ star undergoes first dredge-up when its core mass is between about $0.1\,\rmn{M}_{\sun}$ and $0.2\,\rmn{M}_{\sun}$.
This does not affect our results because we find later that a star must have a core mass greater than about $0.27\,\rmn{M}_{\sun}$ to reach the CSPN region.

\subsection{Mesh spacing function}
The Cambridge stellar-evolution code produces models with meshpoints distributed in a non-Lagrangian self-adaptive mesh according to the method described by \citet{Egg71:351}.
Throughout an evolutionary sequence of models the number of meshpoints is kept constant but the positions evolve according to a chosen mesh spacing function $q$.
Meshpoints are spaced equally in $q$.
We find that the standard mesh spacing function with $200$ meshpoints gives poor resolution, for our purposes, in the hydrogen-burning shell of low-mass RGB star.
This means that when we remove mass from such a star, it undergoes unphysical loops in the HR diagram as its composition profile is distorted by numerical diffusion.
Models computed with a larger number of meshpoints, about $2000$, do not have this problem and converge in terms of the maximum effective temperature and transition time in the post-CE phase at a given core mass.
However, making sequences of models with large numbers of meshpoints takes an inconveniently long time.
To achieve quick and accurate results with fewer meshpoints, we modify the mesh spacing function so that there is a preference for meshpoints around the hydrogen-burning shell.
We achieve this by adding a term to $q$ which varies rapidly in this region,
\begin{equation}
 q = q_0 + C \ln \left[\frac{P + P_{\rmn{low}}}{P + P_{\rmn{high}}}\right],
\end{equation}
where $q_0$ is the standard mesh spacing function, $C$ is a suitably chosen constant and $P$ is the pressure.
This term most rapidly varies when the pressure is between $P_{\rmn{high}}$ and $P_{\rmn{low}}$ and therefore acts to increase resolution there.
In models of low-mass RGB stars we find the pressure changes by about a factor of $10$ in the region of large composition gradient between the core and envelope.
Therefore we choose $P_{\rmn{low}} = 10^{-0.5}P_{\rmn{H}}$ and $P_{\rmn{high}} = 10^{+0.5}P_{\rmn{H}}$, where $P_{\rmn{H}}$ is the pressure in the hydrogen burning shell, approximated as the point at which the hydrogen mass fraction is $0.1$.
We choose the constant $C$ by increasing its value until we find consistent results in sequences with $400$ and $2000$ meshpoints.
We use this modified mesh spacing function with $400$ meshpoints for all models computed in this paper.

\subsection{Thermal non-equilibrium remnants}

\begin{figure}
  \includegraphics[width=84mm]{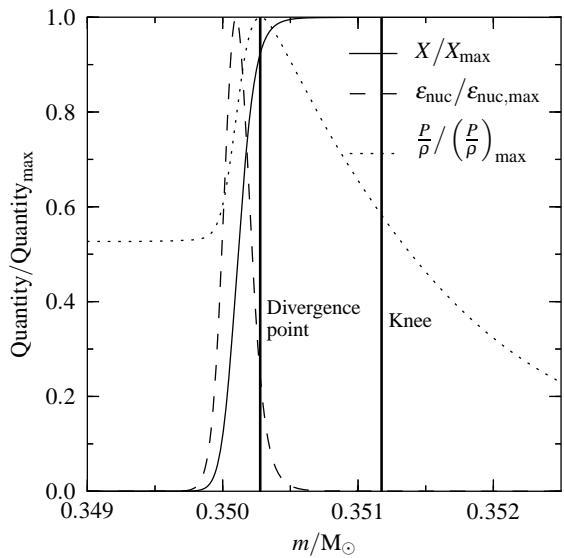}
  \caption{
Structure profiles relevant to prescriptions for the remnant--envelope boundary in a pre-CE red giant of core mass $0.3500\,\rmn{M}_{\sun}$, \mbox{$M_0=1\,\rmn{M}_{\sun}$} and $Z_0=0.02$.
We plot the hydrogen mass fraction $X$, nuclear energy generation rate $\varepsilon_{\rmn{nuc}}$ and the ratio of pressure to density $P/\rho$.
Each quantity is plotted relative to its maximum in the mass range shown.
The divergence point and knee masses are shown.
The base of the convective envelope is at $0.3551\,\rmn{M}_{\sun}$.
The peel-off mass is at about $0.36\,\rmn{M}_{\sun}$.}
  \label{fig:rem-env}
\end{figure}

To make thermal non-equilibrium remnants with masses less than the divergence-point mass is more difficult.
\citet{Iva11:76} applied very large mass-loss rates to produce her models.
We encounter convergence difficulties when we attempt to sustain such high rates to low envelope masses with our code.
This does not matter because the divergence point mass is less than the knee mass, as is demonstrated by the profiles for a $0.3500\,\rmn{M}_{\sun}$ core RGB star in Fig.~\ref{fig:rem-env}.
This is true at all core masses and therefore these stars do not regain thermal equilibrium and so avoid the plateau phase and evolve directly to the cooling track as shown by \citet{Dei70:671}.
In the cases they considered, the evolution to the cooling track is very fast, taking less than about $100\,\rmn{yr}$, so the thermal non-equilibrium remnants of this type do not spend long inside the CSPN region.
We extend this behaviour to various end-of-CE core masses and envelope masses.
However, it is possible that when a remnant of this type reaches the white dwarf cooling track it begins a hydrogen-shell flash or, at high core masses, ignites helium in a late hot flash.
In these cases the same caveats apply as for thermal equilibrium remnants (Section~\ref{sec:flashes}).

\section{Results}
\label{sec:results}
We have argued that the thermal non-equilibrium remnants with masses less than the divergence point mass do not form PNe.
We now analyse our evolutionary models of thermal equilibrium remnants to find the parameters for which a PN is formed and calculate the fading times, according to different prescriptions of post-CE wind mass loss.
We use these results to investigate the importance of PNe formed after CE phases initiated by low-mass red giants.

\subsection{Stars of $\bmath{Z_0=0.02}$, constant mass}\label{sec:Z0.02}

\begin{figure}
  \includegraphics[width=84mm]{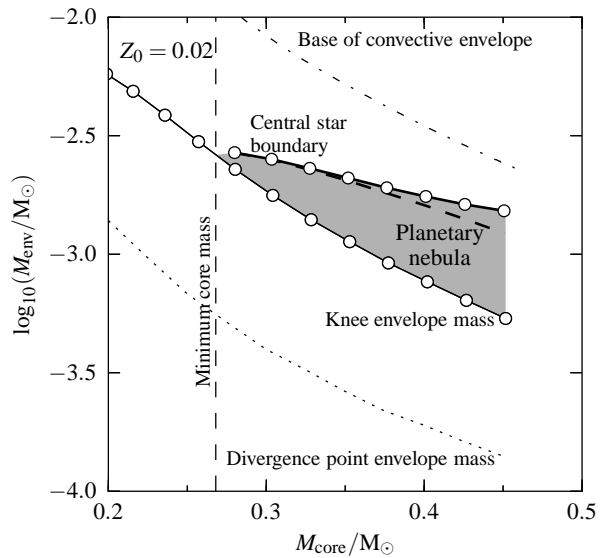}
  \caption{
Combinations of end-of-CE core and envelope mass which form PNe.
Remnants in the shaded region successfully form PNe.
The solid upper boundary to this region represents remnants which reach the critical effective temperature $10^4\,\rmn{yr}$ after the end of the CE phase.
The thick dashed line represents remnants which end the CE phase with the critical effective temperature.
The solid lower boundary to the shaded region represents stars which end the CE phase with the knee envelope mass.
Stars cannot end the CE phase in thermal equilibrium with an envelope mass below the knee envelope mass.
The peel-off mass, the maximum envelope mass for which remnants evolve to become white dwarfs is approximately at the upper boundary of the plot.
Remnants above this line do not represent possible end-of-CE structures.
The uppermost dot--dashed line is the base of the convective envelope.
The vertical dashed line at $0.27\,\rmn{M}_{\sun}$ is the minimum end-of-CE core mass to reach the CSPN region, derived from the data plotted in Fig.~\ref{fig:Mcore-lgTeffmax}.
The lowest dotted line represents the divergence point envelope mass.}
  \label{fig:Mcore-lgMenv1}
\end{figure}

\begin{figure}
  \includegraphics[width=84mm]{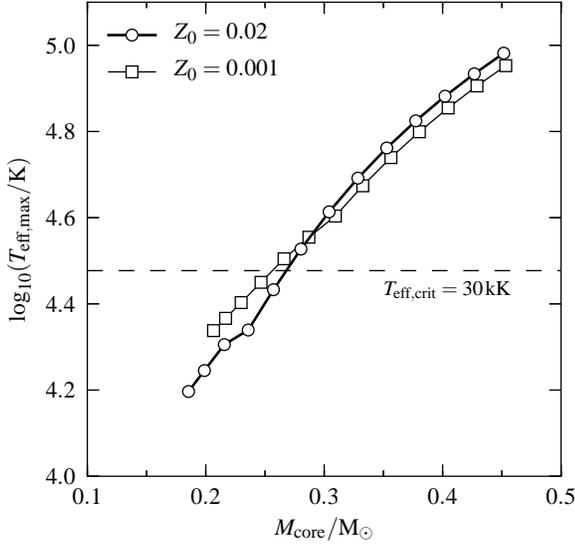}
  \caption{
Maximum effective temperature in the post-CE phase as a function of end-of-CE core mass.
The points are computed from our constant-mass models on the assumption that the core mass is constant during post-CE evolution.
We show results for $Z_0=0.02$ (circles) and $Z_0=0.001$ (squares).
The critical effective temperature, $30\,\rmn{kK}$, is indicated by the horizontal dashed line.
The intersection of the $Z_0=0.02$ line and this horizontal line indicates a remnant of this metallicity must have a core mass greater than about $0.27\,\rmn{M}_{\sun}$ to reach the effective temperature required to form a PN.}
  \label{fig:Mcore-lgTeffmax}
\end{figure}

We first consider $Z_0=0.02$ remnants which evolve with constant total mass.
In Fig.~\ref{fig:Mcore-lgMenv1} we show the core and envelope mass of stars with the critical effective temperature, $30\,\rmn{kK}$ (the thick dashed line).
We also plot the core and envelope mass for stars at the knee stage.
Stars which end the CE phase with envelope masses in the shaded region between these two curves satisfy the conditions for forming a PN because at the end of the CE phase they are in the CSPN region.
Stars which end the CE phase with a core mass less than $0.27\,\rmn{M}_{\sun}$, the vertical dashed line in Fig.~\ref{fig:Mcore-lgMenv1}, do not reach the CSPN region.
This minimum value is computed from the intersection in Fig.~\ref{fig:Mcore-lgTeffmax}, where we have plotted the critical effective temperature and the maximum effective temperature against core mass, assuming that the core mass is approximately constant during the post-CE phase.
\citet{Ibe89:505} found the same minimum core mass to reach the critical effective temperature.

The maximum end-of-CE envelope mass for which a PN is formed is larger than the thick dashed line in Fig.~\ref{fig:Mcore-lgMenv1} because, at higher core masses, it is possible for a remnant to form a PN if it has a transition time less than $10^4\,\rmn{yr}$, the estimated lifetime of the nebula.
This boundary can be computed from our detailed evolutionary models, as we have done for Fig.~\ref{fig:Mcore-lgMenv1}, or it can be derived as follows.
During the plateau phase of a thermal equilibrium remnant, the core mass increases according to
\begin{equation}
  \frac{\rmn{d}M_{\rmn{core}}}{\rmn{d}t} = \frac{L}{X_{\rmn{env}}E_{\rmn{H}}}
\end{equation}
where $X_{\rmn{env}}$ is the hydrogen mass fraction in the envelope and $E_{\rmn{H}} = 6 \times 10^{18}\rmn{\,erg\,g^{-1}}$ is the energy released per unit mass of hydrogen burned \citep{Egg68:387}.
For core masses greater than about $0.2\,\rmn{M}_{\sun}$, $X_{\rmn{env}}=0.677$.
Because, during the plateau phase, the luminosity is approximately constant and a function of the core mass, this equation can be integrated and we can compute the change in envelope mass, $\Delta M_{\rmn{env}}$, to reach the CSPN region within $\Delta t = 10^4\,\rmn{yr}$, assuming it evolves at constant total mass.
We find
\begin{equation}
  \Delta t \approx \frac{\Delta M_{\rmn{env}}X_{\rmn{env}}E_{\rmn{H}}}{L},
\end{equation}
where $\Delta M_{\rmn{env}}$ is computed from the envelope mass at the critical effective temperature.

\begin{figure}
  \includegraphics[width=84mm]{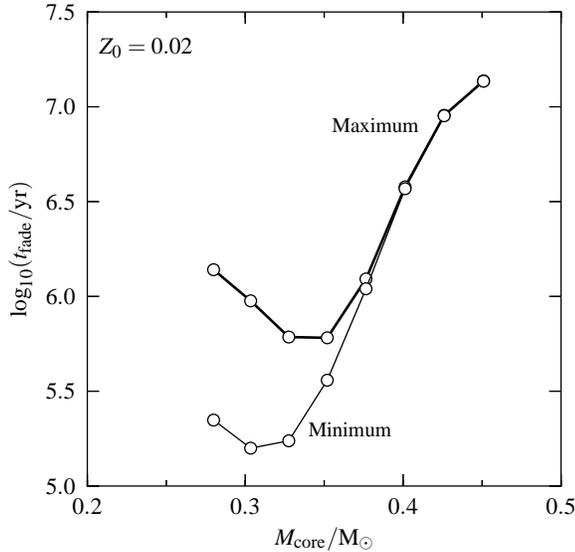}
  \caption{
Maximum and minimum fading times as a function of end-of-CE core mass.
The maximum fading time is the time taken to pass through the PN region of the HR diagram when the remnant ends the CE phase outside this region.
The minimum fading time is the time taken to pass through the CSPN region when the remnant ends the CE phase with the knee envelope mass.}
  \label{fig:Mcore-lgtfade}
\end{figure}

For those remnants which successfully form a PN, we compute the fading time defined in Section~\ref{sec:PNconditions}.
This is necessary to calculate the duration of the PN phase.
This is required to compute the number of PNe expected from this channel (Keller et al.,\ private communication).
In Fig.~\ref{fig:Mcore-lgtfade} we plot the range of fading times as a function of core mass.
The maximum is for a remnant which ends the CE phase with the critical effective temperature.
The minimum is for a remnant which ends the CE phase at the knee.
The fading times are larger than the estimated nebular lifetime of $10^4\,\rmn{yr}$.
To summarize, if a CE remnant must evolve to $30\,\rmn{kK}$ within $10^4\,\rmn{yr}$ after the end of the CE phase to form a PN, a $Z_0=0.02$ star must end the CE phase with a core mass greater than $0.27\,\rmn{M}_{\sun}$.
If it evolves with negligible mass loss, the range of allowed envelope masses is indicated by the shaded region in Fig.~\ref{fig:Mcore-lgMenv1}.
Remnants take between $10^5$ and $10^7\,\rmn{yr}$ to pass through the CSPN region after the end of the CE phase.

\subsection{Stars of $\bmath{Z_0=0.02}$, with post-CE wind mass loss}

\begin{figure}
  \includegraphics[width=84mm]{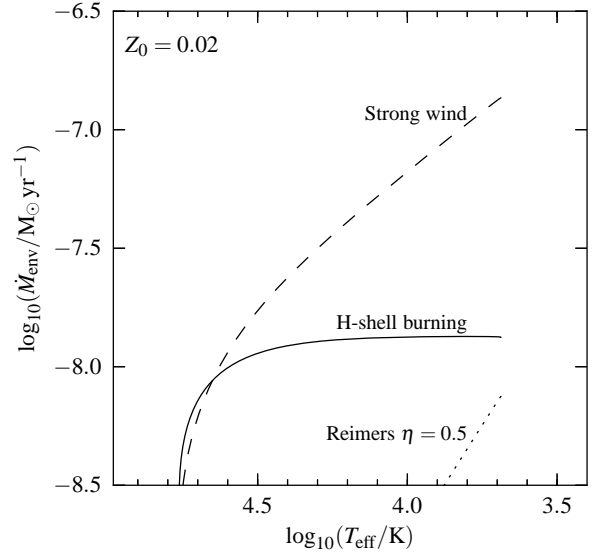}
  \caption{
Rate at which envelope mass is decreased by different processes as a function of effective temperature for a $0.3500\,\rmn{M}_{\sun}$ core mass, $Z_0=0.02$ remnant.
The Reimers' mass-loss formula with $\eta=0.5$ gives a negligible rate of envelope mass decrease compared to that by shell burning.
The strong wind described by equation~(\ref{eq:IT86}) dominates over shell burning at low effective temperature.}
  \label{fig:Mcore-Menvdot}
\end{figure}

If there is significant wind during the post-CE phase then, for a given core mass, the range of end-of-CE envelope masses for which PNe are formed is increased because the remnant evolves more rapidly to high effective temperature.
The post-CE plateau evolution is well modelled by a sequence of models of approximately constant core mass and decreasing envelope mass.
Because of this, we can think of the envelope mass as dictating the effective temperature and therefore the rate at which the envelope mass decreases as dictating the rate of effective temperature evolution.
A wind decreases the envelope mass in addition to the decrease from hydrogen shell-burning and therefore increases the rate of evolution of effective temperature.

In Fig.~\ref{fig:Mcore-Menvdot} we show possible contributions to the rate at which envelope mass is decreased as a function of effective temperature for a $0.3500\,\rmn{M}_{\sun}$ core mass $Z_0=0.02$ remnant.
We show the rate at which the envelope mass is decreased by shell burning and two possibilities for the uncertain wind mass-loss rate.
One possibility would be an unjustified application of Reimers' formula (equation~\ref{eq:Reimersrate}) with the standard value of $\eta=0.5$.
We disregard this because the rate at which the envelope mass is decreased is negligible compared to shell burning.
Another possibility is the extrapolation of the mass-loss rate formula computed by \citet{Vin02:553} for horizontal branch stars.
Again, we find that this gives a negligible rate of envelope mass decrease compared to shell burning at all core masses.
Finally, we consider an extreme, strong wind, upper limit as suggested by \citet{Ibe86:742}.
They estimated a maximum mass-loss rate by assuming that the photons leaving the surface of the star lose all their momentum to the kinetic energy of the wind so $\dot{M}v_{\rmn{esc}} = L / c$, where $v_{\rmn{esc}}=\sqrt{2GM/R}$ is the escape speed and $c$ is the speed of light.
Assuming the total mass is approximately constant during the plateau phase, this gives
\begin{equation}\label{eq:IT86}
  \frac{\rmn{d}M_{\rmn{env}}}{\rmn{d}t} = -5.6 \times 10^{-11} \left(\frac{L}{\rmn{L}_{\sun}}\right)\left(\frac{R}{\rmn{R}_{\sun}}\right)^{1/2}\,\rmn{M}_{\sun}\,\rmn{yr}^{-1}.
\end{equation}
The rate at which the envelope mass decreases according to this equation is also shown Fig.~\ref{fig:Mcore-Menvdot} where it is referred to as a strong wind.
In this extreme case and at low effective temperatures, the wind is the dominant process by which the envelope mass decreases.

\begin{figure}
  \includegraphics[width=84mm]{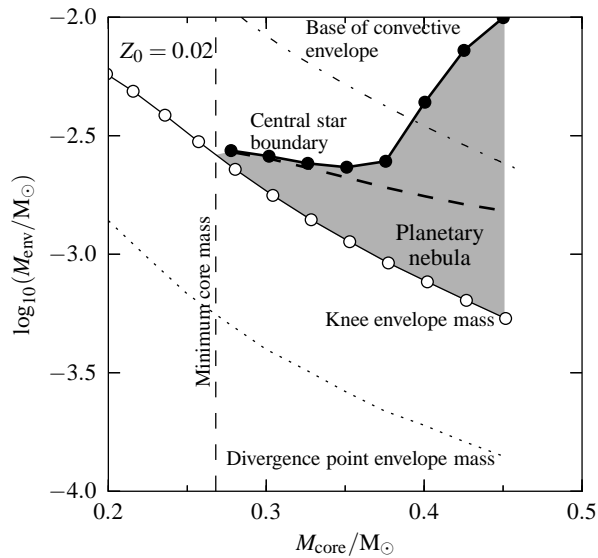}
  \caption{
Combinations of end-of-CE core and envelope mass which form PNe.
As Fig.~\ref{fig:Mcore-lgMenv1}, except here we include the effect of wind during the post-CE phase on the region for which PNe are formed.
The upper boundary for remnants evolving at constant mass is the thick dashed line.}
  \label{fig:Mcore-lgMenv3}
\end{figure}

We computed models with this mass-loss rate and remnants with the peel-off mass ($10^{-2}\,\rmn{M}_{\sun}$).
In Fig.~\ref{fig:Mcore-lgMenv3} we show how this changes the range of end-of-CE envelope masses for which PNe are formed at each core mass.
There is a larger range of envelope masses than when there is no wind and significantly so for core masses greater than about $0.3\,\rmn{M}_{\sun}$.
At lower core masses the rate at which the envelope mass is decreased by wind becomes comparable to the shell burning rate.
For core masses greater than about $0.4\,\rmn{M}_{\sun}$, evolution with such a strong wind implies that a star could end the CE phase with an mass near the peel-off mass and still form a PN.
This is not realistic, because we do not expect this mass-loss rate to be valid for stars at low effective temperatures with convective envelopes.
It is valid only for masses below the peel-off mass.
A strong wind does not have a large effect on the fading times because the mass-loss rate decreases as the remnant reaches the knee stage.
The fading times are still greater than $10^5\,\rmn{yr}$, as for post-CE stars evolving at constant mass.

\subsection{Prescriptions for the end-of-CE structure}
\label{sec:prescriptions}
\begin{figure}
  \includegraphics[width=84mm]{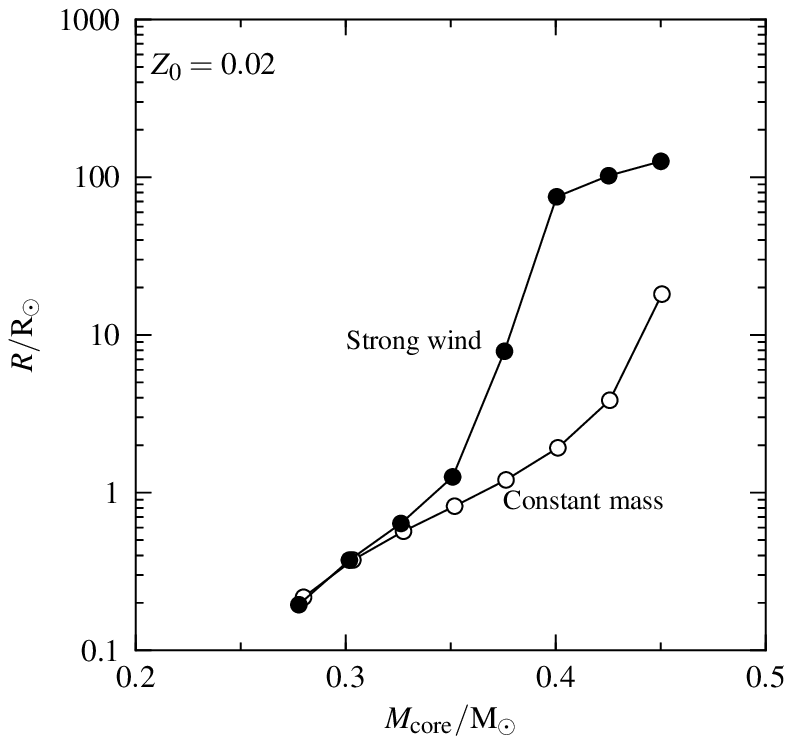}
  \caption{
Maximum end-of-CE radius as a function of core mass for $Z_0=0.02$ remnants which successfully form PNe.
The maximum radius is shown for the cases of post-CE evolution at constant mass (open circles) and with a strong wind (filled circles, equation~\ref{eq:IT86}).
For a remnant to successfully form a PN it must end the CE phase with radius below the curves.}
  \label{fig:Mcore-lgRatTeffcrit}
\end{figure}

\begin{figure}
  \includegraphics[width=84mm]{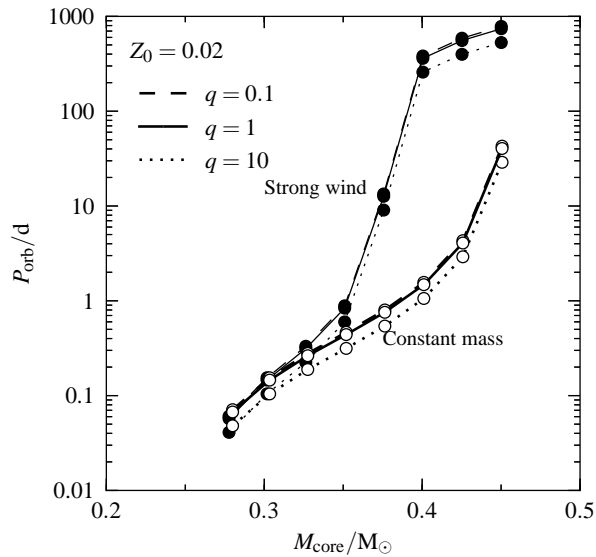}
  \caption{
Maximum end-of-CE orbital period as a function of core mass for $Z_0=0.02$ remnants which successfully form PNe.
The maximum period is shown for the cases of post-CE evolution at constant mass (open circles) and with a strong wind (filled circles, equation~\ref{eq:IT86}).
For a remnant to successfully form a PN it must end the CE phase with orbital period below the curves, which are shown for three different post-CE mass ratios, \mbox{$q=M_{\rmn{core}}/(\text{mass of companion})$}.}
  \label{fig:Mcore-lgPorb}
\end{figure}

We now transform our results to a more useful form by assuming that the remnant just fills its Roche lobe at the end of the CE phase, as suggested by \citet{Ibe89:505}.
If this is the case then a $Z_0=0.02$ system with a given core mass must end the CE phase with a Roche lobe radius less than those plotted in Fig.~\ref{fig:Mcore-lgRatTeffcrit}.
This relation implies a core mass--maximum orbital period relation with a weak dependence on the mass ratio (Fig.~\ref{fig:Mcore-lgPorb}).
Many of the observed post-CE systems in the SDSS sample described by, for example, \citet{Zor11:L3} satisfy these conditions and therefore we expect they underwent a phase as a PN.
Synthetic populations also contain systems which satisfy these conditions \citep*[e.g.,][]{Dav10:179}.

In simple prescriptions of CE phases initiated by low-mass red giants the end-of-CE envelope mass is a function of core mass only.
We can use Fig.~\ref{fig:Mcore-lgMenv1} to determine the range of core masses for which PNe are formed in each prescription.
If the remnant--envelope boundary is at $X=0.1$ then remnants have zero end-of-CE envelope mass.
If the remnant--envelope boundary is at $X=0$ then remnants have negative envelope mass with our definition.
Both of these prescriptions imply end-of-CE masses below the divergence point mass.
We have argued that such remnants do not undergo a plateau phase and therefore do not spend a sufficiently long time in the CSPN region for the formation of a PN.
If the remnant--envelope boundary is at the base of the convective envelope then remnants do not form PNe for any core mass because they evolve too slowly to reach the CSPN region before the nebula has dispersed.

\subsection{Stars of $\bmath{Z_0=0.001}$}\label{sec:Z0.001}
\begin{figure}
  \includegraphics[width=84mm]{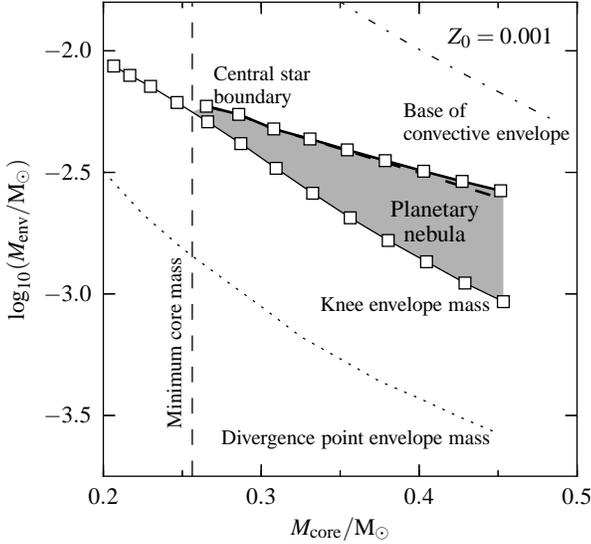}
  \caption{
Combinations of end-of-CE core and envelope mass which form PNe.
As Fig.~\ref{fig:Mcore-lgMenv1} but for systems of $Z_0=0.001$.
Envelope masses are larger compared to $Z_0=0.02$ remnants of the same core mass.}
  \label{fig:Z0.001-Mcore-lgMenv-1}
\end{figure}

\begin{figure}
  \includegraphics[width=84mm]{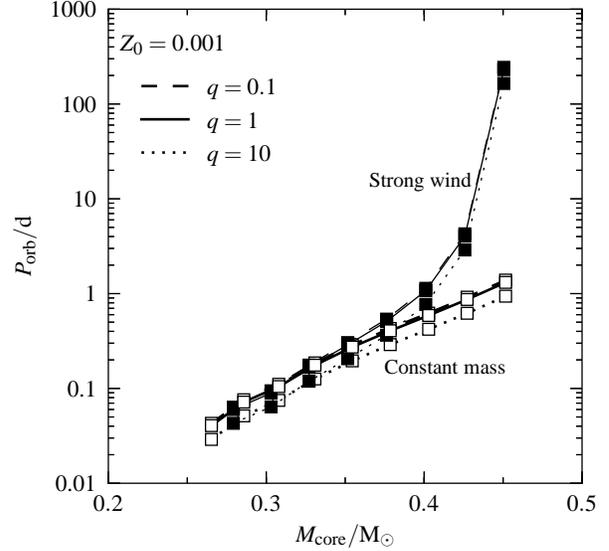}
  \caption{
Maximum end-of-CE orbital period as a function of core mass for $Z_0=0.001$ remnants which successfully form PNe.
The maximum period is shown for the cases of post-CE evolution at constant mass (open squares) and with a strong wind (filled squares, equation~\ref{eq:IT86}).
For a remnant to successfully form a PN it must end the CE phase with orbital period below the curves, which are shown for three different post-CE mass ratios, \mbox{$q=M_{\rmn{core}}/(\text{mass of companion})$}.
Maximum orbital periods at the same core mass are shorter than those for $Z_0=0.02$ remnants.
}
  \label{fig:Z0.001-Mcore-lgPorb}
\end{figure}

We repeated our analysis for $Z_0=0.001$ stars.
At this metallicity, remnants must have a core mass greater than about $0.26\,\rmn{M}_{\sun}$ to reach the CSPN region (Fig.~\ref{fig:Mcore-lgTeffmax}).
We show the equivalent of Fig.~\ref{fig:Mcore-lgMenv1}, the end-of-CE core mass--envelope mass plane, in Fig.~\ref{fig:Z0.001-Mcore-lgMenv-1}.
Envelope masses are larger than those of $Z_0=0.02$ remnants of the same core mass.
The maximum end-of-CE orbital periods at a given core mass are lower than for $Z_0=0.02$, as shown in Fig.~\ref{fig:Z0.001-Mcore-lgPorb}.

\section{Discussion}
\label{sec:discussion}
We now discuss the implications of these results for the question of the importance of PNe formed after CE phases initiated by low-mass red giants.

\subsection{Observations of nebulae}
We have not found any reported PNe that can be unambiguously identified as post-RGB systems.
There are several systems which had been identified as non post-AGB and therefore possibly post-RGB.
However, it has since been argued that some of these nebulae are ionized regions of the interstellar medium rather than ejected envelopes.
Notable examples are the star PHL~932, the central star of EGB~5 \citep{Men88:L25}, and the nebulae Sh~2-174, K~2-2 \citep{Nap99:101} and DeHt~5 \citep{Bar01:1149}.
\citet{Fre10:203} showed that the velocity of PHL~932 is inconsistent with the velocity of the nebula surrounding it and similar arguments have been made for the cases of Sh~2-174 \citep{Fre08:0, Fre10:129}, K~2-2 and DeHt~5 \citep{DeM13:2118}.
EGB~5 is another suggested example of this type of system \citep{Fre08:0, Fre10:129}, but further work is needed.
This system is important now that a companion to the hot star has been discovered \citep{Gei11:L16}.
There are also some possibly post-RGB systems in which the nebula does appear to have been ejected from the hot central star, but in these cases alternative interpretations have not been ruled out.
The atmospheric parameters HDW~11 are consistent with a post-RGB interpretation \citep{Nap99:101}.
Another possible example is VW Pyx in K~1-2 for which the parameters of the central system are uncertain \citep*{Ex03:1349}.

It is therefore not clear whether there are any post-RGB/CE PNe and, if there are any, how frequent they are.
The situation is further confused by the possibility that some of the systems which are identified as post-AGB PNe have not been considered as post-RGB.
Because of this, we discuss confusing and distinguishing the two evolutionary states.

\subsection{Recognizing post-RGB systems}
The main difference between a PN formed after a CE phase and a PN formed from an isolated star is that in the post-CE case, the central ionizing source has a companion in a short period orbit.
But what properties can be used to distinguish between PNe formed after CE phases initiated by RGB and AGB stars?
We focus on low-mass stars.

The properties of the central system are the stellar properties of the components and the orbital properties.
In a PN formed by a low-mass RGB star, the central ionizing source is a pre- helium-core white dwarf with mass less than about $0.5\,\rmn{M}_{\sun}$.
We have shown that PNe are only expected to have central stars with masses greater than about $0.26\,\rmn{M}_{\sun}$ in post-RGB/CE systems.
In our models we find this means it has a luminosity of less than about $3.16 \times 10^3\,\rmn{L}_{\sun}$.
Post-AGB stars are expected to be pre-carbon--oxygen white dwarfs with larger masses and luminosities than this.
Although the lowest mass post-AGB stars could be confused with post-RGB stars on the basis of evolutionary tracks in terms of surface gravity, effective temperature and luminosity alone, especially when we allow for uncertainties in the distance to systems.

The properties of the nebula include composition, size, expansion speed, morphology, electron density and luminosity.
To a first approximation the composition of the nebula is that of the envelope of the giant at the moment of ejection.
The abundances of various elements are different for low-mass RGB and AGB stars because of nucleosynthesis and first-, second- and third-dredge-up events.
A post-AGB nebula would be expected to show more evidence of nuclear processing than a post-RGB nebula.
In the absence of detailed models of CE evolution it is difficult to say what would be the differences in size, expansion speed and morphology, electron density and luminosity.
We would need the initial conditions of the ejected circumstellar matter, which can only come from a detailed model.

\subsection{Observations of systems without nebulae}
We have found that a pre-white dwarf plateau phase is only expected after a CE phase if the envelope mass is larger than the knee envelope mass.
Thus the observation of a post-RGB star in the high gravity plateau phase, with or without an ionized nebula, would allow us to distinguish between the two prescriptions we have discussed.
Again, we are not aware of any known systems of this type.
Such systems would be similar to the known pre-low mass white dwarf systems which have been found.
These include HD 18812 \citep{Heb03:L477}, 1SWASP J024743.37−251549.2 \citep{Max11:1156}, KOI-74, KOI-81, KIC 10657664 and KOI 1224 \citep{Bre12:115}.
In these systems the core mass is thought to be less than $0.25\,\rmn{M}_{\sun}$.
More interesting here are post-RGB/CE systems with larger core masses than this.
These would be expected to contribute to the population of hot subdwarfs (sdOs) in short-period binary systems.
These sdO stars are more usually explained as descendants of post-CE sdB stars, that is they are the remnants of stars burning helium in the core with a thin hydrogen envelope \citep{Str07:269}.
One possible example of a system in a post-CE plateau phase is AA Dor.
In this system the hot component could be a post-RGB or post-extreme horizontal branch star based on its surface gravity and effective temperature \citep{Kle11:L7}.

\section{Conclusions}
\label{sec:conclusions}
We have considered the contribution of common-envelope phases initiated by low-mass red giants to the planetary nebula population.
Although this subset is a large fraction of the common-envelope phases which end with ejection of the envelope, it has sometimes been assumed that the remnants evolve too slowly to ionize the nebula before it has dispersed.
We have more carefully investigated this assumption by considering the possible structures at the end of the common-envelope phase and their evolution in the theoretical Hertzsprung--Russell diagram.
We find that a planetary nebula is not expected if the remnant envelope is removed to the divergence point defined by \citet{Iva11:76}.
Planetary nebulae are expected if the remnants end the phase in thermal equilibrium with envelope masses such that they fill their Roche lobes \citep{Ibe89:505}.
We have derived a relation between the maximum Roche lobe radius and core mass for which planetary nebulae are formed in this prescription, both with and without a strong post-common envelope wind.
We have extended the work of \citet{Ibe89:505} to lower metallicity systems and the case of a strong wind.
Lower metallicity remnants must emerge from the common-envelope in shorter period orbits for planetary nebulae to form at the same core mass.

Although we have yet to find any unambiguously identified post-red giant branch planetary nebulae, we should not discount this as a possible evolutionary interpretation.
Remnants of post-asymptotic giant branch and post-red giant branch stars can be distinguished by the mass of the ionizing component, the abundances in the nebula and the photospheres of the hot remnants.
Nebulae of this type would be useful to learn about common-envelope evolution and the formation of planetary nebula.
In future work we shall use these results to construct synthetic populations of planetary nebulae and assess the observational evidence that they may be formed after common-envelope phases initiated by low-mass red giants.

\section*{Acknowledgements}
PDH thanks the Science and Technology Facilities Council, STFC, for his studentship. 
CAT thanks Churchill College for his fellowship and all it entails.
RGI thanks the Alexander von Humboldt Foundation for funding his position in Bonn.
We thank the referee, Noam Soker, for his helpful comments.

\bibliographystyle{mn2e_arxiv}
\bibliography{paper_arxiv}

\label{lastpage}
\end{document}